\documentstyle[prl,aps,epsf,twocolumn]{revtex}
\begin{document}
\title{
L\'evy Distribution of Single Molecule
Line Shape Cumulants in Low Temperature Glass
 }
\author{E. Barkai$^a$, R. Silbey$^a$ and G. Zumofen$^b$\\
$^a$ Department of Chemistry and Center for Material Science and Engineering,
Massachusetts Institute of Technology, Cambridge, MA 02139.\\
$^b$ Physical Chemistry Laboratory, Swiss Federal Institute of Technology, ETH-Z, CH-8092 Z$\ddot{u}$rich. \\
}
\date{\today}
\maketitle

\begin{abstract}

{\bf Abstract} 

We investigate the distribution of single molecule line shape
 cumulants, $\kappa_1,\kappa_2,\cdots$,
in low temperature glasses
based on the sudden jump, standard tunneling model. 
We find that the cumulants are described 
by L\'evy stable laws, thus generalized central limit theorem
is applicable for this problem.

\end{abstract}

Pacs: 05.40.-a, 05.40.Fb, 61.43.FS, 78.66.Jg \\



%

 Recent experimental advances \cite{MO} have made it possible to measure
the spectral line shape of a single molecule 
(SM) embedded in a condensed phase. 
Because each molecule is in a unique static and dynamic environment,
the line shapes of chemically identical SMs vary from molecule 
to molecule \cite{Tittel}.
In this way, the dynamic properties of the host are encoded
in the distribution of single molecule spectral line shapes 
\cite{MO,Tittel,Reilly2,Zumofen,Geva,Pfluegl,Orrit,Barkai1,Barkai2}.
We examine the statistical properties of the line shapes
and show how these are related to the underlying microscopic  
dynamical events occurring in the condensed phase.

 We use the
Geva--Skinner \cite{Geva}
 model for the SM line shape in a low temperature
glass 
based on the sudden jump picture of 
Kubo and Anderson \cite{AK,Reilly}.
In this model, a random distribution of low--density (and
non--interacting) dynamical defects 
[e.g., spins or two level systems (TLS)] interacts with
the molecule via {\em long range 
interaction} 
(e.g., dipolar). 
We show that L\'evy statistics fully characterizes the
properties of the SM spectral line  both in the {\em fast}  and {\em slow
modulation limits}, while far from these limits L\'evy statistics
describes the mean and variance of the line shape. 
We then compare our
analytical results, derived in the slow modulation limit, with results 
obtained from numerical simulation.
The good agreement indicates that the slow modulation limit is correct
for the parameter set relevant to experiment.


 L\'evy stable distributions 
serve as
a natural generalization of the normal Gaussian distribution.
The importance of the Gaussian in statistical physics
stems from the central limit theorem.  L\'evy stable laws are used
when analyzing sums of the type $\sum x_i$, with 
$\{ x_i \}$ being independent identically distributed
 random variables
characterized by a diverging  variance.
In this case the ordinary Gaussian central limit theorem must be replaced
with the generalized central limit theorem.
With this generalization, 
L\'evy stable probability densities, $L_{\gamma,\eta}(x)$, replace
the  Gaussian of the standard central
limit theorem. 
Khintchine and L\'evy found
that stable characteristic functions,
$\hat{L}_{\gamma,\eta}(k)$,
 are of the form \cite{Feller}
\begin{equation}
\ln\left[ \hat{L}_{\gamma,\eta} \left( k \right) \right] = i \mu k 
- z_{\gamma}|k|^{\gamma} 
\left[ 1 - i \eta { k \over |k|} 
\tan\left( { \pi \gamma \over 2}\right)\right]
\label{eq1}
\end{equation}
for $0< \gamma\le 2$ (for the case $\eta \ne 0$, $\gamma=1$ see
\cite{Feller}). Four parameters are needed for a full description of 
a stable law. The constant $\gamma$ is called the characteristic exponent, 
the parameter $\mu$ is a location parameter which is unimportant in
the present case, $z_{\gamma} > 0$ is a scale parameter and $-1\le \eta \le 1$
is the index of symmetry. When $\eta=0$ $(\eta=\pm 1)$
the stable density $L_{\gamma,\eta}(x)$ is symmetrical (one sided).
L\'evy statistics is known  to describe
several long range interaction systems in 
diverse fields such as  astronomy \cite{Feller}, turbulence \cite{Turbo}
and spin glass  \cite{Jap}. 
Stoneham's theory \cite{Stoneham}
of inhomogeneous line broadening in defected crystal, is based on long-range forces
and parts of it can be
interpreted in terms of L\'evy stable laws \cite{Barkai2}. 
Stoneham's approach \cite{Stoneham} is inherently static, while the SM
line shape model  
considers both dynamical and static contributions from
the defects.

%

An important issue is the slow and fast modulation limits
\cite{Pfluegl,Reilly}. 
Briefly, the  fast (slow) modulation limit is valid if important
contributions to the line shape are from TLSs which satisfy
$\nu \ll K$ $(\nu \gg K)$, where 
$\nu$ is the frequency shift of the SM due to 
SM--TLS
interaction
and $K$ is the 
transition rate of the flipping TLS
(see details below).
In the fast modulation limit,
all (or most) lines are Lorentzian 
with a width 
that varies from one molecule to the other. For this case, the (L\'evy)
distribution
of {\em line widths} fully characterizes the statistical properties
of the lines. 
The second, more complicated, case corresponds
to the slow modulation
limit. Then the SM line is typically 
composed of several peaks
(splitting) and is not described well by a Lorentzian. 
If a SM shows splitting, one
can investigate the validity of the
standard tunneling model of glass \cite{AHV} in a direct way, 
since the splitting of
a line is directly associated with SM--TLS interaction \cite{Orrit}. 
As mentioned, we demonstrate the existence
of a slow modulation limit in SM--glass system.

 
 Following \cite{Geva}
we assume a SM coupled to non identical independent TLSs
at distances ${\bf r}$ in dimension $d$.
Each TLS is characterized by its asymmetry variable $A$
and tunneling element $J$.
The energy of the TLS is $E=\sqrt{A^2 + J^2}$.
The TLSs are coupled to phonons
or other thermal excitations
such that the state of the TLS changes with time. The state of
the $n$ th TLS is described by an occupation parameter,
$\xi_n(t)$, that is equal $0$ or $1$ if the TLS is in its
ground or excited state respectively. The probability for finding the TLS
in its upper $\xi=1$ state, 
$p$, is given 
by the standard Boltzmann form
$p=1/\{ 1+ \exp[ E/(k_b T)]\}$. 
The transitions
between the ground and excited state are described by the
up and down transition rates $K_u,K_d$, which  are related to
each other by the standard detailed balance condition.

 The excitation of the $n$ th TLS shifts the SM's
transition frequency by $\nu_n$. Thus, the SM's
transition frequency is
\begin{equation}
\omega(t) = \omega_0 + \sum_{n=1}^{N_{\rm act}} \xi_n \left( t \right) \nu_n,
\label{eq2}
\end{equation}
where $N_{\rm act}$ is the number of active TLSs in the system
(see details below) and 
$\omega_0$ is the bare transition frequency that differs
from one molecule to the other depending on the
local static disorder.
We consider a wide class of frequency perturbations 
\begin{equation}
\nu = 2 \pi \alpha \Psi\left( \Omega\right) f\left( A, J \right)
{ 1 \over r^{\delta}},
\label{eq3}
\end{equation}
$\alpha$ is a coupling constant with units
$\left[ \mbox{Hz}\ \mbox{nm}^{\delta}\right]$,
$\Psi\left(\Omega \right)$ is a dimensionless
function of order unity, $\Omega$ is a vector of angles
determined by the orientations of the TLS and molecule
(in some simple cases $\Omega$ depends on polar angles
only),
$f(A,J)\ge 0$  is a dimensionless function of the internal
degrees of freedom of the fluctuating TLS,  
$\delta$
is the interaction exponent. 
 The line shape of the SM is given by
the complex Laplace transform of the 
relaxation function
\begin{equation}
I_{SM}\left( \omega \right)={1 \over \pi} \mbox{Re}
\left[  \int_0^\infty dt e^{i \omega t} 
\Pi_{n=1}^{N_{\rm act}}\Phi_n\left( t \right) \right]
\label{eq4}
\end{equation}
provided that the natural life time of the SM excited state is long.
The relaxation function  of a single TLS was evaluated \cite{Reilly} based
on methods developed in \cite{AK}
\begin{equation}
\Phi\left( t \right) = e^{-\left( \Xi+ i p\nu\right) t}
\left[\cosh\left( \Omega t \right) + {\Xi\over \Omega}
\sinh\left( \Omega t \right) \right]
\label{eq5}
\end{equation}
with
%
$\Omega = [K^2/ 4 - \nu^2 / 4 - i \left( p - 1/2 
 \right) \nu K]^{1/2}$,
%
$\Xi = { K \over 2} - i \left( p - {1 \over 2} \right) \nu$
%
and $K=K_u + K_d$.
For a bath of TLSs the line shape, Eq. (\ref{eq4}), 
is a formidable function of the random TLS
parameters $(r,\Omega,A,J)$ as well as the system
parameters $( \alpha , T, \mbox{etc})$.
In the fast modulation limit
$K\gg |\nu|$, one finds a simpler behavior: all lines are Lorentzian
with half width
\begin{equation}
\tilde{\Gamma}= \sum_n^{N_{act}} p_n(1-p_n)\nu_n^2/K_n
\label{eqGamma}
\end{equation}
which varies from one molecule to the other.
Eq. (\ref{eqGamma}) shows the well known
phenomena of motional narrowing.
In the slow modulation limit $K \ll |\nu|$
one finds  
$ \Phi(t)=1 - p + pe^{- i \nu t}$ implying that the line shape
of a molecule coupled to a single TLS 
is composed of two delta peaks, the line shape 
of a molecule coupled to
two TLSs is composed of four delta peaks, etc (splitting).

 The spectral line is characterized by its cumulants
$\kappa_j$ $(j=1,2,\cdots)$ that vary from one molecule to the
other, and we investigate the cumulant probability density
$P(\kappa_j)$.
 We have derived the cumulants of the SM line shape, 
and the first four cumulants are
presented in Table 1 \cite{remark}. 
We observe that cumulants of order 
$j\le 2$ are real while generally cumulants of
order $j>2$ are complex,
implying that  the moments of the line shape
diverge when $j>2$. The summation, $\sum_n$, in Table $1$
is over the active TLSs, namely those TLSs which flip
on the time scale of observation $\tau$ (i.e., $K_n > 1/ \tau$).
 We consider the slow modulation limit, soon to be justified,
which means that we consider the case $K_n \ll \nu_n$.
To investigate this limit we set $K_n=0$ in Table $1$,
then all the cumulants
are real and are rewritten as $\kappa_j=\sum_n H_{j n} \nu_n^j$,
where $H_j$ are functions of $p$ only
and $H_1=p$, $H_2=p(1-p)$, $H_3=p(1-p)(2p-1)$ etc.
Note that for $\kappa_1$ and $\kappa_2$
no approximation 
is made.

\begin{tabular}{|l|l|}
\hline
j & $\kappa_j$  \\
$ $ & $$ \\
$1$ &$ \sum_n p_n \nu_n $  \\
$ $ & $$ \\
$2$ &$  \sum_n p_n\left(1 - p_n\right)  \nu_n^2  $   \\
$ $ & $$ \\
$3$ &$ \sum_n p_n\left(1- p_n\right)\left( 2 p_n - 1\right) \nu_n^3
+i \sum_n p_n(1-p_n)K_n \nu_n^2 $   \\
$ $ & $$ \\
$4$ &$ \sum_n p_n\left(-1+ p_n\right)
\left[ K_n^2 + \nu_n^2\left(-1 + 6p_n - 6 p_n^2\right) \right]\nu_n^2 -$\\
$ \ $ &$
2 i \sum_n  K_n \left( - 1 + p_n \right) p_n \left( - 1 + 2 p_n\right) \nu_n^3
$   \\
$ $ & $$ \\
\hline
\end{tabular}
$$\mbox{Table  1: Cumulants $\kappa_j$ of the SM line shape} $$

Let 
$\left\langle \cdot \right\rangle_{r \Omega A J}$  denote an averaging
over
the random TLS parameters.
The characteristic function
of the $j$ cumulant can be written in a form
$$ \left\langle \exp \left( i k \kappa_j\right)\right \rangle_{r \Omega A J} = $$
\begin{equation}
\exp\left[ - \rho_{\mbox{eff}} \left\langle \int d \Omega  \int_0^{\infty}
{d (r^d) \over d}\left( 1 - e^{i k B_j/ r^{\delta j }}\right)\right\rangle_{AJ}\right],
\label{eq8}
\end{equation} 
$ \rho_{\mbox{eff}}$ is the density of the active TLS and 
$B_j=(2 \pi \alpha)^j \Psi^j \left( \Omega \right) f^j \left(A,J\right)H_j$.
To derive Eq. (\ref{eq8}) we have
used the assumption of independent
TLSs uniformly distributed in the system.
 For odd $j$  cumulants we find
\begin{equation}
\left\langle \exp\left(i k \kappa_j \right) \right\rangle_{r \Omega A J}=
\hat{L}_{\gamma,0}( k ),
\label{eq9}
\end{equation}
with characteristic exponent $\gamma=d/(\delta j )$ and the scale parameter
\begin{equation}
z_{\gamma}= \rho_{\mbox{eff}} \left( 2 \pi \alpha\right)^{{d\over \delta}}
\left\langle f^{d/\delta}\left( A , J \right) |H_j|^{\gamma} \right\rangle_{AJ}
c_{\gamma} \int d \Omega |\Psi^j\left( \Omega\right)|^{\gamma}
\label{eq10}
\end{equation}
with $c_{\gamma} = \cos\left( \gamma \pi/2\right)\Gamma(1 - \gamma)$,
$c_1=\pi/2$.
Eq.  (\ref{eq9}) shows that
odd cumulants are described by symmetrical L\'evy stable density, i.e.,
$P(\kappa_j)=L_{\gamma,0}(\kappa_j)$. 
Two conditions must be satisfied for such
a behavior,  $0 <\gamma < 2$ and 
$\int d \Omega \sin\left[ \Psi^j(\Omega)\right]=0$.  The latter condition
gives the symmetry condition, $\eta=0$,  which means that negative
and positive contributions to $\kappa_j$ are equally probable.

 For even cumulants and $0< \gamma < 1$ we find
\begin{equation}
\left\langle \exp\left(i k \kappa_j \right) \right\rangle_{r \Omega A J}=
\hat{L}_{\gamma,\eta}(k)
\label{eq11}
\end{equation}
with a scale parameter Eq. (\ref{eq10})
and with L\'evy index of symmetry 
\begin{equation}
\eta= { \left\langle f^{j \gamma} \left( A, J \right) |H_j|^{\gamma} { H_j \over |H_j|} \right\rangle_{AJ} \over
\left\langle f^{j \gamma} \left( A, J \right) |H_j|^{\gamma} \right\rangle_{AJ}}.
\label{eq12}
\end{equation}
Eq. (\ref{eq11}) implies that even cumulants are distributed
according to $P(\kappa_j)=L_{\gamma,\eta}(\kappa_j)$.
We note that the asymmetrical L\'evy functions,
with $\eta\ne \pm1,0$, only rarely find
their applications in the literature.
The characteristic exponent $\gamma$ depends only on the general features
of the model (namely on $d$ and $\delta$). In contrast the
L\'evy index of symmetry  $\eta$ depends on the details
of the model and on system
parameters $(T,\mbox{etc})$. For $j=2$ we
have $H_j=|H_j|$ and then $\eta=1$ so
the L\'evy density is one sided, as is expected since
$\kappa_2 > 0$.

 As mentioned, in the fast modulation limit, 
the random line width in Eq. (\ref{eqGamma}) characterizes the
statistical properties of the spectral lines. 
Using the approach in Eqs. (\ref{eq8}-\ref{eq10})
one can show that $P(\tilde{\Gamma})=L_{d/(2 \delta),1}(\tilde{\Gamma})$
with the scale parameter $z_{d/(2 \delta)}$ given by Eq. (\ref{eq10})
with $j=2$ and $H_2=p(1-p)/K$.
 
 In what follows we exhibit our results and compare to 
simulations based on the standard
tunneling model of low temperature glass \cite{AHV}. We use system parameters
given by Geva and  Skinner \cite{Geva} to model
terrylene in polystryrene. 
The SM-TLS interaction is dipolar,
hence $\delta =3$, and we consider spatial dimension $d=3$.
The distribution
of the asymmetry parameter and tunneling element 
is $P(A)P(J)=N^{-1} J^{-1}$ for 
$2.8\times 10^{-7}{\rm K} < J <18 {\rm K}$ and
$ 0< A < 17{\rm K}$, $N$ denoting a normalization constant. 
We use $f(A,J)=A/E$ and define a  TLS to be active if
$K> 1/\tau$, $\tau=120$ sec is the time of experimental observation.
In this way the averaging $\left\langle \cdots \right\rangle_{AJ}$ becomes
$\tau$ independent. 
The rate of the TLS is given by 
$K = c J^2 E \coth\left({\beta E_n/2} \right)$ 
and $c=3.9 \times 10^{8} {\rm K}^{-3}$Hz 
is the TLS phonon coupling constant.
Additional system parameters are the coupling constant 
$\alpha=3.75\times 10^{11}$ $\mbox{nm}^{3}$Hz and the
TLS density $1.15\times 10^{-2}$ nm$^{-3}$.
According to
Eqs. (\ref{eq9})-(\ref{eq12}), only the scale parameter $z_{\gamma}$
depends on 
the orientation of the TLS and SM, through $\Psi(\Omega)$.
It is therefore reasonable to assume simple forms for $\Psi(\Omega)$.
We consider two examples,  
model $1$  ($\mbox{M}1$) for which  $\Psi(\Omega)$
is replaced with a two state variable (i.e., a spin model)
$\Psi=1$ or $\Psi=-1$ with equal probabilities of occurrence
and model $2$ ($\mbox{M}2$)
$\Psi(\Omega)=\cos(\theta)$, with $\theta$, the standard polar
coordinate, distributed uniformly.
With these definitions we calculate the symmetry index
$\eta$ and the scaling parameter $z_{\gamma}$ and compare between
the theory and numerical simulation.


\begin{figure}[htb]
\epsfxsize=20\baselineskip
\centerline{\vbox{
      \epsffile{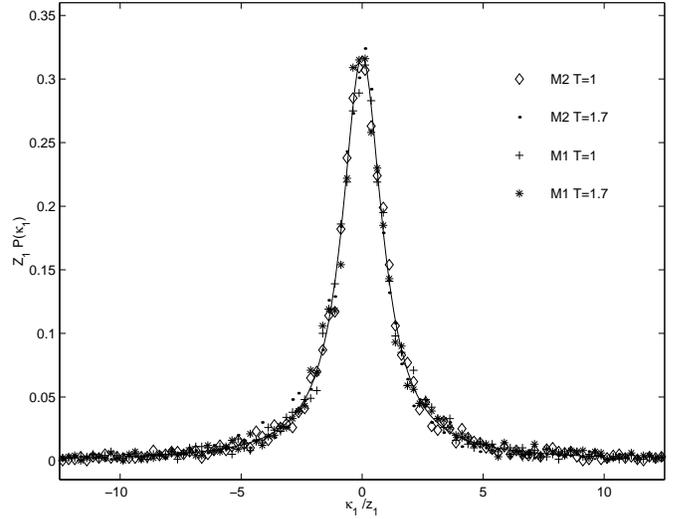}  }}
\caption {
Scaled probability density of first cumulant
$P(\kappa_1)z_1$ versus  $\kappa_1/z_1$.
Symbols are the simulation results obtained from $4000$
molecules for different cases indicated in the figure.
The theory, plotted as a solid curve, 
predicts a Lorentzian density $P(\kappa_1)=L_{1,0}(\kappa_1)$
with a scaling parameter $z_1$ which varies from one set of data
to the other.
}
\label{fig1}
\end{figure}

\begin{figure}[htb]
\epsfxsize=20\baselineskip
\centerline{\vbox{
      \epsffile{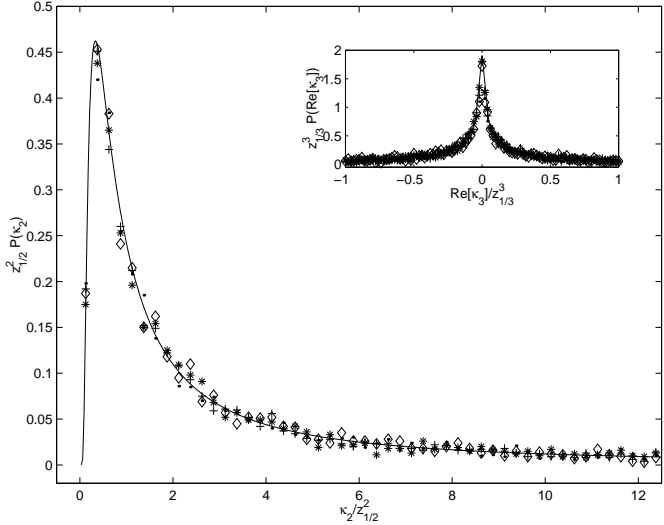}  }}
\caption {
Same as Fig. (\protect{\ref{fig1}}) 
for the second cumulant. We show $P(\kappa_2)z_{1/2}^2$ versus
$\kappa_2/z_{1/2}^2$. The solid curve is Smirnov's density 
$L_{1/2,1}(\kappa_2)$. 
In the inset we show
the same as Fig. (\protect{\ref{fig1}}) for $\mbox{Re}[\kappa_3]]$,
the solid curve is
L\'evy density $L_{1/3,0}(\mbox{Re}[\kappa_{1/3}])$.
}
\label{fig2}
\end{figure}

 We consider the first two cumulants $\kappa_1$ and $\kappa_2$
(i.e, the line shape mean and variance). 
Since $d=\delta$ we find $P(\kappa_1)=L_{1,0}(\kappa_1)$
which is the Lorentzian density,
and $P(\kappa_2) = L_{1/2,1}(\kappa_2)$ which is 
Smirnov's
density. We have considered two temperatures for the two models
M1 and M2. As shown in Fig. \ref{fig1} and \ref{fig2},
a scaling behavior is observed and all data collapse on the
L\'evy densities $L_{1,0}(\kappa_1)$ and $L_{1/2,1}(\kappa_2)$ respectively.
In Fig. \ref{fig1} and \ref{fig2} we have rejected TLSs within a sphere
of radius $r_{\rm min}=1$ nm,
demonstrating that our results are not sensitive to a short cutoff.
Also shown in the inset of Fig. \ref{fig2} is 
$P(\mbox{Re}[\kappa_3])$ which is distributed
according to $L_{1/3,0}(\mbox{Re}[\kappa_3])$ and
a scale parameter $z_{1/3}$ given in 
Eq. (\ref{eq10}).
The L\'evy behavior of $\kappa_1,\kappa_2$ and
$\mbox{Re}[\kappa_3]$
holds generally and is not limited to the slow modulation limit
since these random variables do not depend explicitly on the 
rates $K$. 

\begin{figure}[htb]
\epsfxsize=20\baselineskip
\centerline{\vbox{
      \epsffile{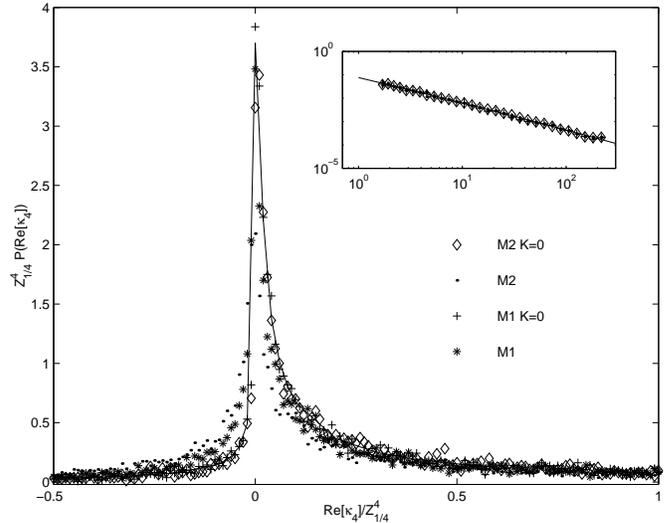}  }}
\caption {
Same as Fig. 
(\protect{\ref{fig1}}) 
for $Re[\kappa_4]$ and for temperature $T=1.7 {\rm K}$.
The symbols are the simulation results obtained for four different
cases as indicated in the figure and for $12000$
molecules.
The solid curve is $L_{1/4,\eta}[\mbox{Re}(\kappa_4)]$ with
an index of asymmetry $\eta=0.6104$.
In the inset we show the power law
tail of the scaled probability density on a log-log plot.
}
\label{fig4}
\end{figure}

 Consider the distribution of $\mbox{Re}[\kappa_4]$,
which in the slow modulation limit is distributed according
to $L_{1/4,\eta}(\mbox{Re}[\kappa_4])$, 
Eq. (\ref{eq11}). The question
remains if such a slow modulation limit is valid for the standard
tunneling model parameters we are considering. The slow modulation limit
is expected to work when $K\ll |\nu|$. For large enough $r$
this inequality will fail; however,
depending on system parameters, we expect that
contributions from TLS situated far from the SM
are negligible. We also note
that according to the standard tunneling model the rates
$K$ are distributed over a broad range, albeit with
finite cutoffs that insure that the averaged rate 
is finite.
To check if the slow modulation limit
is compatible with the standard tunneling model approach,
we compare 
our slow modulation results 
with those obtained by simulation 
in Fig. \ref{fig4}. 
We also show
simulation results in which all rates are set to zero ($K=0$).
For model
$\mbox{M}1$, we find that the deviation between simulation
and theory is small so the assumption of slow modulation
limit is justified. For model $\mbox{M} 2$, we see slightly
larger deviations between the theory and numerical results,
due to
the angular dependence of model $\mbox{M} 2$,
$\Psi(\Omega) = \cos(\theta)$, which reduces the typical frequency
shift $|\nu|$ compared to model $\mbox{M}1$. We conclude that
the present theory 
can be used as a criterion for the validity 
of the slow modulation limit. 

%

 Depending on system parameters, L\'evy statistics
may become sensitive to the finite cutoff  $r_{\rm min}$,
Physically, the cutoff can be important since the
power low interaction is not supposed to work well
for short distances \cite{Pfluegl}.
Our results were derived for $r_{\rm min}=0$,
while for small though finite $r_{\rm min}$ 
one can find intermittency behavior, i.e.,
the ratio $\langle \kappa_2^2 \rangle / \langle \kappa_2\rangle^2$
(as well as similar dimensionless ratios) is very
large. 
When $r_{\rm min}$ is large one finds a Gaussian behavior.
The phenomena of intermittency in the context of a reaction
of a SM in a random environment was investigated
in \cite{Wang}.
Generally high order cumulants are more sensitive to finite
cutoff and for results in Fig (\ref{fig4}) $r_{\rm min}=0$
was chosen to see the proper decay laws in the
wing.

 To conclude, we showed that the generalized central limit theorem
can be used to analyze distribution of cumulants of SM
line shapes in glass. 
We note that besides cumulants,
L\'evy statistics can be used to analyze other
statistical properties of SMs in disordered media \cite{Barkai2}. 

{\bf Acknowledgment} EB thanks the ETH and Prof. Wild for
their hospitality. This work was supported by the NSF.


\begin{thebibliography}{99}
\bibitem{remark} In Table 1 we use $\omega_0=0$, for $\omega_0\ne 0$
we have $\kappa_1=\sum_n p_n \nu_n + \omega_0$ while the higher order 
cumulants are $\omega_0$ independent. The distribution of $\omega_0$ depends
on the static disorder \cite{Stoneham} while we are considering
the dynamic disorder due to the flipping TLSs.
\end{thebibliography}
\end{document}